\documentstyle[11pt]{article}
\def\be{\begin{equation}}
\def\ee{\end{equation}}
\def\ba{\begin{array}{rcl}}
\def\ea{\end{array}}
\def\le#1{\label{eq:#1}}
\def\eq#1{(\ref{eq:#1})}

\def\a{\alpha}        \def\b{\beta}        \def\g{\gamma}
\def\d{\delta}             \def\ve{\varepsilon} 
\def\m{\mu}           \def\n{\nu}           
        \def\S{\Sigma}
       \def\vphi{\varphi}
        \def\o{\omega}                    

\def\cL{{\cal L}}     \def\cO{{\cal O}}      
\def\cH{{\cal H}}     \def\cP{{\cal P}}

\font\bfi=cmmib10 
\def\mb#1{\hbox{\bfi #1}}
\def\wbar#1{\hskip1pt\overline{\phantom J}\hskip-9pt #1{}}
\def\what#1{\widehat #1{}}
\def\wtilde #1{\widetilde #1{}}
\def\fr#1#2{\hbox{$\frac{#1}{#2}$} }            
\def\subsub#1{\par {\bf #1} }
\def\vs{\vspace{6pt}}
\def\vss{\vspace{12pt}}

\null
\begin{document}

\begin{flushright}
IF--15/97 \\
December 1997
\end{flushright}
\vss

\begin{center}
{\LARGE \bf Conservation Laws in Poincar\'e} \\  
\vs
{\LARGE\bf Gauge Theory}\footnote{Talk presented at the workshop {\em Gauge
Theories of Gravitation\/}, Jadwisin, Poland, September (1997) } \\
\vss\vss\vss
{Milutin Blagojevi\'c\footnote{e--mail: mb@casandra.phy.bg.ac.yu }} \\
\vs
{Institute of Physics, 11001 Belgrade, P.O.Box 57, Yugoslavia }           
\end{center}
\vs
\begin{abstract}
Basic features of the conservation laws in the Hamiltonian approach to
the Poincar\'e gauge theory are presented. 
     It is shown that the Hamiltonian is given as a linear combination
of ten first class constraints. The Poisson bracket algebra of these
constraints is used to construct the gauge generators.
By assuming that the asymprotic symmetry is the global Poincar\'e
symmetry, we derived the improved form of the asymptotic generators,
and discussed the related conservation laws of energy, momentum, etc.   
\end{abstract}
\vs

Among various attempts to overcome the problem of quantization of
the general relativity, gauge theories of gravity are especially
attractive, as they are based on the concept of gauge symmetry which has
been very successfull in describing other fundamental interactions in
nature. The importance of the Poincar\'e symmetry in particle physics
leads one to consider the Poincar\'e gauge theory (PGT) as a natural
framework for the description of the gravitational phenomena~\cite{1}.

In this paper we shall 
{$~a)$} present the Hamiltonian structure of the general PGT~\cite{2,3}, 
{~$b)$} construct the gauge generators \cite{4,5,6}, and 
{~$c)$} clarify the relation between gauge symmetries and
conservation laws, in case of asymptotically flat spacetimes~\cite{7}.  

\pagebreak

\section{Hamiltonian dynamics} 

The Hamiltonian analysis of PGT leads to a simple form of
the gravitational Hamiltonian, and yields a clear picture of the dynamical 
structure \cite{3}.

Basic gravitational variables in PGT are tetrad $b^i{_\m}$  and
Lorentz connection $A^{ij}{_\m}$, and the
corresponding field strenghts are torsion and curvature: 
$T^i{_{\m\n}}=\partial_\m b^i{_\n}+A^i{_{s\m}}b^s{_\n} 
                               -(\m\leftrightarrow\n)$, 
$\,R^{ij}{_{\m\n}}=\partial_\m A^{ij}{_\n} +A^i{_{s\m}}A^{sj}{_\n} 
                               -(\m \leftrightarrow\n)$.
The geometry of PGT is defined by the Riemann--Cartan spacetime $U_4$,
with the general Lagrangian 
$\wtilde\cL=b\cL_G\bigl(R^{ij}{_{kl}},T^i{_{kl}}\bigr) 
                            +b\cL_M\bigl(\Psi ,\nabla_k\Psi \bigr)$,
where $\Psi$ are matter fields and $\nabla_k\Psi$ is the covariant
derivative. The gravitational Lagrangian which is at most
quadratic in field strenghts, i.e. of $R+T^2+R^2$ type, depends
on ten parameters (assuming parity invariance). 

\subsub{Constraints.}  
The momentum variables $(\pi_k{^\m},\pi_{ij}{^\m},\pi)$, corresponding
to $(b^k{_\m},A^{ij}{_\m},\Psi)$, are obtained frpm $\tilde\cL$ in
the usual way. 
Due to the fact that $T^i{_{\m\n}}$ and $R^{ij}{_{\m\n}}$ are defined
through the antisymmetric derivatives of $b^k{_\m}$ and $A^{ij}{_\m}$,
respectivly, they do not involve velocities of $b^k{_0}$ and
$A^{ij}{_0}$. As a consequence, one immediately obtains the following
set of the so--called {\it sure\/} primary constraints:
\be
\phi_k{^0}\equiv\pi_k{^0}\approx 0 \, ,\qquad 
   \phi_{ij}{^0}\equiv\pi_{ij}{^0}\approx 0 \, .              \le{1}
\ee
These constraints are allways present, independently of the values of
parameters in $\tilde\cL$. Depending of a specific form of the
Lagrangian, one may also have additional primary constraints in
the theory. 

The {\it canonical Hamiltonian\/} has the form
${\cH}_c={\cH}_M+{\cH}_G$, where 
${\cH}_M=\pi\Psi_{,0}-\wtilde{\cL}_M$, 
${\cH}_G=\pi_k{^\a}b^k{_{\a ,0}}
  + {\fr 12}\pi_{ij}{^\a}A^{ij}{_{\a ,0}}-\wtilde{\cL}_G$.
The {\it total Hamiltonian\/} is  
\be
{\cH}_T={\cH}_c + u^k{_0}\phi_k{^0}
        + {\fr 12}u^{ij}{_0}\phi_{ij}{^0} +(u\cdot\phi ) \, , \le{2}
\ee
where $\phi$ denotes all additional primary constraints,
if they exist (if--con\-straints), and $H_T=\int d^3x\cH_T$.  

The evaluation of the consistency conditions of the primary
constraints, $\dot\pi_k{^0}=\{\pi_k{^0},H_T\}\approx 0$ and   
$\dot\pi_{ij}{^0}=\{\pi_{ij}{^0},H_T\}\approx 0$, 
is essentially simplified if we previously find out the
dependence of $\cH_T$ on the unphysical variables $b^k{_0}$ and
$A^{ij}{_0}$. We shall show that ${\cH}_c$ is linear in $b^k{_0}$ and
$A^{ij}{_0}$, 
\be
{\cH}_c=b^k{_0}{\cH}_k -{\fr 12}A^{ij}{_0}{\cH}_{ij}+\partial_\a D^\a\, ,
                                                              \le{3}
\ee
where $\partial_\a D^\a$ is a three--divergence term, while possible
extra primary constraints $\phi$ are independent of $b^k{_0}$ and
$A^{ij}{_0}$. Consequently, the consistency conditions of the primary
constraints will result in the {\it secondary\/} constraints:  
\be
{\cH}_k\approx 0\, ,\qquad {\cH}_{ij}\approx 0 \, .           \le{4}
\ee

The linearity of $\cH_c$ in $b^k{_0}$ and $A^{ij}{_0}$ is closely
related to the so--called (3+1) decomposition of spacetime.  
If $\mb{n}$ is the unit normal to the hypersurface 
$\S_0: x^0=\hbox{const.}$, the four vectors $\{\mb{n},\mb{e}_\a\}$
define the so--called ADM basis. 
Introducing the projectors on $\mb{n}$ and $\S_0$,
$(P_\bot )_k^l=n_kn^l \, ,\, (P_\parallel )_k^l=\d_k^l-n_kn^l$,
we can express any vector in terms of its parallel and orthogonal
components: $V_k=n_k V_\bot + V_{\bar k}$, where 
$V_{\bar{k}}\equiv (V_\parallel )_k = (P_\parallel )_k^l\, V_l$,  
$V_\bot =V^kn_k$. An analogous decomposition can be defined for any
tensor.

The decomposition of $\mb{e}_0$ in the ADM basis yields
$\mb{e}_0=N\mb{n} + N^\a\mb{e}_\a$, where $N$ and $N^\a$ are called
lapse and shift functions, respectivly.
By using the fact that $N$ and $N^\a$ are linear in $b^k{_0}$,
$N= n_kb^k{_0}$, $N^\a =h_{\bar{k}}{^\a}b^k{_0}$,
the canonical Hamiltonian \eq{3} can be easily brought into an
equivalent form: 
\be
{\cH}_c = N{\cH}_\bot + N^\a{\cH}_\a 
      -{\fr 12}A^{ij}{_0}{\cH}_{ij} + \partial_\a D^\a \, ,  \le{5}
\ee
where $\cH_\bot =n^k\cH_k$, $\cH_\a =b^k{_\a}\cH_k$.
\subsub{Matter Hamiltonian.} Let us now turn to the proof of 
\eq{5} for the matter Hamiltonian. First, we decompose $\nabla_k\Psi$
into the orthogonal and parallel components, 
$$
\nabla_k\Psi =n_k\nabla_\bot\Psi +\nabla_{\bar k}\Psi\equiv n_kh_\bot{^\m}
 \nabla_\m\Psi + h_{\bar k}{^\a}\nabla_\a\Psi \, .            
$$
Replacing this into $\cL_M$ leads to 
$\cL_M=\wbar\cL_M(\Psi ,\nabla_{\bar{k}}\Psi ;\nabla_\bot\Psi ,n^k)$,
where complete dependence on velocities and unphysical variables 
$(b^k{_0},A^{ij}{_0})$ is contained in $\nabla_\bot\Psi$. Second, since
$b=\det (b^k{_\m}) = NJ$, where $J$ does not depend on $b^k{_0}$, 
the expression for $\pi$ can be written as  
$$
\pi \equiv {\partial (b\cL_M)\over\partial\Psi_{,0}}
    = J{\partial\wbar\cL_M\over \partial \nabla_\bot\Psi} \, .
$$
Finally, using the relation
$\nabla_0\Psi\equiv N\nabla_\bot\Psi +N^\a\nabla_\a\Psi
   =\Psi_{,0}+ {\fr 12}A^{ij}{_0}\S_{ij}\Psi$
to express the velocities $\Psi_{,0}$, the canonical Hamiltonian for
matter fields takes the form \eq{5}, where
\be
\ba
&&\cH_\a^M=\pi\nabla_\a\Psi \, ,\qquad {\cH}_{ij}^M=\pi\S_{ij}\Psi\, ,\cr
&&\cH_\bot^M=\pi\nabla_\bot\Psi -J\wbar\cL_M \, ,\qquad D^M_\a=0\, .\le{6}
\ea
\ee
Expressions for ${\cH}_\a^M$ and ${\cH}_{ij}^M$ are independent of
unphysical variables. They do not depend on the
specific form of $\cL_M$, but only on the transformation properties of
fields, and are called {\it kinematical\/} parts of the Hamiltonian. 
The term ${\cH}_\bot^M$ is {\it dynamical\/}, as it depends on the
choice of $\cL_M$. After eliminating $\nabla_\bot\Psi$ with
the help of the relation defining $\pi$, one finds that 
${\cH}_\bot^M$  does not depend on unphysical variables:  
${\cH}_\bot^M={\cH}_\bot^M(\Psi ,\nabla_{\bar{k}}\Psi ;\pi /J,n^k)$.

Additional primary constraints, if they exist, are also independent of
unphysical variables. 
\subsub{Gravitational Hamiltonian.} Construction of the gravitational
Ha\-mil\-to\-ni\-an can be performed in a very similar way, the role of
$\nabla_k\Psi$ being taken over by $T^i{_{km}}$ and $R^{ij}{_{km}}$. In
the first step we decompose the torsion and the curvature, in last
two indices, into the orthogonal and parallel components.
The parallel components $T^i{_{\bar k\bar m}}$ and 
$R^{ij}{_{\bar k\bar m}}$ are independent of velocities and unphysical
variables. The replacement in the gravitational Lagrangian yields 
$\cL_G=\wbar\cL_G(T^i{_{\bar k\bar m}},R^{ij}{_{\bar k\bar m}};
                T^i{_{\bot\bar k}},R^{ij}{_{\bot\bar k}},n^k)$.
The relations defining gravitational momenta take the form
$$
{\hat\pi}_i{^{\bar k}}=J{\partial\wbar\cL_G\over
                   \partial T^i{_{\bot\bar k}}}\, ,\qquad 
{\hat\pi}_{ij}{^{\bar k}}=J{\partial\wbar\cL_G\over
                      \partial R^{ij}{_{\bot\bar k}}}\, ,
$$
where ${\hat\pi}_i{^{\bar k}}\equiv\pi_i{^\a}b^k{_\a}$ and 
${\hat\pi}_{ij}{^{\bar k}}\equiv\pi_{ij}{^\a}b^k{_\a}$  are ``parallel"
gravitational momenta. The velocities $b^i{_{\a,0}}$ and
$A^{ij}{_{\a,0}}$ can be calculated from the definitions of 
$T^i{_{0\a}}$ and $R^{ij}{_{0\a}}$. After a simple algebra the
canonical Hamiltonian takes the form \eq{5}, where 
\be\ba
&&{\cH}_{ij}^G=2\pi_{[i}{^\a}b_{j]\a}+2\pi_{k[i}{^\a}A^k{_{j]\a}}
              +\partial_\a\pi_{ij}{^\a} \, ,\cr
&&{\cH}_\a^G=\pi_i{^\b}T^i{_{\a\b}}+ {\fr 12}\pi_{ij}{^\b}R^{ij}{_{\a\b}}
            -b^k{_{\a}}\nabla_\b\pi_k{^\b} \, ,\cr
&&{\cH}_\bot^G= \bigl( {\hat\pi}_i{^{\bar m}}T^i{_{\bot\bar m}}
+ {\fr 12}{\hat\pi}_{ij}{^{\bar m}}R^{ij}{_{\bot\bar m}}-J\wbar\cL_G\bigr)
              -n^k\nabla_\b\pi_k{^\b} \, ,\cr
&&D_G^\a =b^i{_0}\pi_i{^\a}+ {\fr 12}A^{ij}{_0}\pi_{ij}{^\a}\, .  \le{7}
\ea\ee
The expressions $T^i{_{\bot\bar m}}$ and $R^{ij}{_{\bot\bar m}}$ 
in ${\cH}_\bot^G$ should be eliminated with the help of the equations
defining momenta ${\hat\pi}_i{^{\bar m}}$ and 
${\hat\pi}_{ij}{^{\bar m}}$.
\subsub{Consistency of the theory.} 
The fact that $\cH_c$ is linear in unphysical variables implies
the existence of the secondary constraints:  $\cH_\bot\approx 0$,
$\cH_\a\approx 0$ and $\cH_{ij}\approx 0$.
By working out the constraint algebra we shall see that these
constraints are FC. As a consequence, the consistency conditions of
the secondary constraints will be {\it automatically\/} satisfied.

\section{Gauge symmetries} 

The correct definition of gauge generators enables one
to clarify the relationship between gauge symmetries and conservation
laws.   

\subsub{Constraint algebra.} 
An explicite knowledge of the algebra of constraints is necessary for
the investigation of the consistency of the theory, as well as for the
construction of the gauge generators~\cite{4}. 

If extra constraints are not present in the theory, one can show that
the Poisson bracket algebra of the secondary constraints takes the form
\be\ba
&&\{\cH_{ij},\cH'_{kl}\} ={\fr 12}f_{ij}{^{mn}}{_{kl}}\cH_{mn}\d \, ,
 \qquad \{\cH_{ij},\cH'_\a \} = 0 \, ,\cr
&&\{\cH_\a ,\cH'_\b \} =\bigl(\cH_\a '\partial_\b +\cH_\b\partial_\a 
  - {\fr 12}R^{ij}{_{\a\b}}\cH_{ij}\bigr)\d\, ,\cr
&&\{\cH_{ij},\cH'_\bot \} =0 \, ,
 \qquad \{\cH_\a ,\cH'_\bot \} =\bigl(\cH_\bot\partial_\a 
    -{\fr 12}R^{ij}{_{\a\bot}}\cH_{ij}\bigr)\d \, ,\cr
&&\{\cH_\bot ,\cH'_\bot \} =-\bigl({^3}g^{\a\b}\cH_\a +
  {^3}g^{\prime\a\b}\cH_\a'\bigr)\partial_\b\d\, .                \le{8}  
\ea\ee

In the presence of extra constraints the whole analysis becomes much
more involved, but the results are essentially the same:   
~{$a)$} the dynamical Hamiltonian $\cH_\bot$ goes over into a
redefined expression $\wbar\cH_\bot$, that includes the contributions
of all primary second class constraints;
~{$b)$} the Poisson bracket algebra may contain primary FC terms
$(C_{PFC})$. 
Therefore, consistency conditions of the
secondary constraints are automatically satisfied. 

\subsub{Gauge generators.} 
In PGT, the gauge generator has the form 
$G=\dot\ve(t)G_1+\ve(t)G_0$, where $G_0, G_1$
are phase space functions satisfying the conditions~\cite{5} 
\begin{eqnarray}
G_1&=&C_{PFC} \, , \nonumber \\
G_0+\{ G_1,H_T\}&=&C_{PFC}\, , \nonumber \\
\{ G_0,H_T\}&=&C_{PFC} \, . \nonumber
\end{eqnarray}

It is clear that the construction of the gauge generator demands the 
knowledge of the algebra of constraints. Since the 
Poincar\'e gauge symmetry is always present, independently of a
specific form of the action, one naturally expects that all essential
features of the gauge generator can be obtained by
considering the simple case of the theory without extra constraints.
In that case the primary constraints $\pi_k{^0}$ and $\pi_{ij}{^0}$
are FC, and the Poincar\'e gauge generator takes the form \cite{6}    
\be
\tilde G=\int d^3x \bigl[ \dot\xi^\m\bigl( b^k{_\m}\pi_k{^0}
  + {\fr 12}A^{ij}{_\m}\pi_{ij}{^0} \bigr) +\xi^\m\cP_\m 
  + {\fr 12}\dot\o^{ij}\pi_{ij}{^0}+ {\fr 12}\o^{ij}S_{ij} \bigr]\, ,
                                                              \le{9}
\ee
where
\begin{eqnarray}
&&\cP_\m =b^k{_\m}\cH_k- {\fr 12}A^{ij}{_\m}\cH_{ij}
  +b^k{_{0,\m}}\pi_k{^0}+ {\fr 12}A^{ij}{_{0,\m}} \pi_{ij}{^0} \, ,
                                                      \nonumber \\
&&S_{ij}=-\cH_{ij}+2b_{[i0}\pi_{j]}{^0}+2A^s{_{[i0}}\pi_{sj]}{^0} \, . 
                                                      \nonumber
\end{eqnarray}
Note that $\cP_0 =  \cH_T - \partial_\a D^\a$,
since $\,\dot b^k{_0}=u^k{_0}\,$, $\,\dot A^{ij}{_0}= u^{ij}{_0}$, on
shell. 

The action of the gauge generator on the fields 
$(\Psi,b^k{_\m},A^{ij}{_\m})$ produces the correct Poincar\'e gauge
transformations. These transformations are symmetry transformations of
the action not only when extra constraints are absent, but also in the
general case. This fact leads to the conclusion that the expression
\eq{9} is the correct generator of the Poincar\'e gauge symmetry for any
choice of the parameters of the theory.  

\section{Conservation laws}  

Now, we are going to consider one of the most important problems of the
classical theory of gravity --- the definition of the gravitational
energy, and other conserved quantities~\cite{7}.  
\subsub{The asymptotic symmetry.} We assumme that the symmetry of the 
$U_4$ theory in the asymptotic region is the global Poincar\'e
symmetry. The global Poincar\'e transformations
can be obtained from the gauge transformations by the following
replacement of parameters: 
$$
\o^{ij}(x)\to -\o^{ij} \, ,\qquad
\xi^\m (x)\to -\o^\m{_\n}x^\n -\ve^\n\equiv -\xi^\m \, ,   
$$
where $\o^{ij}$ and $\ve^\n$ are constants, 
$\o^\m{_\n}=\d^\m_i\o^{ij}\eta_{j\n}$. 
The related generator can be obtained from the gauge generator \eq{9}
in the same manner, leading to 
\be
G= {\fr 12}\o^{ij}M_{ij}-\ve^\nu P_\nu \, ,                   \le{10}
\ee
where
\begin{eqnarray}
&&P_{\m}=\int d^3x \cP_\m ,                        \nonumber \\
&&M_{\a\b}=\int d^3x\bigl(x_\a\cP_\b -x_\b\cP_\a -S_{\a\b}\bigr) \, ,
                                                    \nonumber \\
&&M_{0\b}=\int d^3x\bigl(x_0\cP_\b -x_\b\cP_0-S_{0\b}+b^k{_\b}\pi_k{^0} 
   + {\fr 12}A^{ij}{_\b}\pi_{ij} {^0}\bigr) \, .   \nonumber         
\end{eqnarray}

Since the generators act on basic dynamical variables via Poisson
brackets, they are required to have {\it well defined functional
derivatives\/}. As this is not always the case with the generator
\eq{10}, we shall try  to improve its form so as to obtain the
expression with well defined functional derivatives. The first step in
that direction is to define precisely the phase space in which the
generator \eq{10} acts.   
\subsub{The phase space.} The choice of asymptotics will become more
clear if we first express the asymptotic structure of spacetime in
certain geometric terms.
Here we shall be concerned with {\it isolated} physical systems,
characterized by matter fields that decrease sufficiently fast at
large distances, so that their contribution to surface integrals 
vanishes. The spacetime outside an isolated system is said to be {\it
asymptotically flat\/} if the following two conditions are satisfied:

$(a)$  $g_{\m\n}=\eta_{\m\n}+\cO_1$, where $\eta_{\m\n}$ is the
Minkowskian metric, $\cO_n$ decreases like $r^{-n}$ or faster for large
$r$, and $r^2=(x^1)^2+(x^2)^2+(x^3)^2$.   

$(b)$ $R^{ij}{_{\m\nu}}=\cO_{2+\a}$ $(\a >0)$ (the absolute parallelism
for large $r$).  

The second condition can be easily satisfied by demanding
$A^{ij}{_\m}=\cO_{1+\a}$.  In the Einstein--Cartan (EC) theory the
connection behaves as  $\partial g_{\m\n}$, so that $A=\cO_2$. The same
law holds in the general $U_4$ theory when the field $A$ is massive,
while massless $A$ can have a slower decrease. We shall study here, for
simplicity, the EC theory, i.e. we shall assume that 
\be
b^k{_\m}=\d^k_\m +\cO_1 \, ,\qquad A^{ij}{_\m}=\cO_2 \, .     \le{11}
\ee
To ensure the global Poincar\'e invariance of these conditions
we demand $b^k{_{\m ,\n}}=\cO_2$, $A^{ij}{_{\m ,\n}}=\cO_3$, etc.

The asymptotic behaviour of {\it momenta\/} is determined
by requiring $p -{\partial\cL /\partial{\dot q}} =\what\cO$,
where $\what\cO$ denotes a term that decreases sufficiently fast.
From the definitions of the gravitational momenta in EC theory one
obtains    
\be
\pi_k{^0}, \pi_{ij}{^0}=\what\cO\, ,\qquad\pi_k{^\a} = \what\cO \, ,
 \quad  \pi_{ij}{^\a}=-4aJn_{[i}h_{j]}{^\a}+\what\cO \, .     \le{12}  
\ee
Similar arguments lead to the consistent determination of the
asymptotic behaviour of the Hamiltonian multipliers.

\subsub{Improving the Poincar\'e generators.}  
The generators act on dynamical variables via Poisson brackets,
defined in terms of functional derivatives. A functional
$F[\vphi ,\pi ]=\int d^3x 
   f\bigl(\vphi,\partial_\m\vphi,\pi,\partial_\n\pi\bigr)$
has well defined functional derivatives if its variation can be written
as  $\d F=\int d^3x \bigl[ A\d\vphi+B\d\pi\bigr]$, 
where terms $\d\varphi_{,\m}$ and $\d\pi_{,\m}$ are {\it absent\/}. 

The variation of the spatial translation generator has the form  
\be\ba
& &\d P_\a =-\d E_\a +R \, ,\cr
& &E_\a\equiv\oint ds_\g\bigl(\pi_{ij}{^\b}A^{ij}{_{[\a}}\d_{\b ]}{^\g}
                                               \bigr) \, ,   \le{13}
\ea\ee
where the integration domain is the boundary of the
three--di\-men\-si\-o\-nal space, and $R$ denotes regular terms, not
containing $\d\vphi_{,\m}$, $\d\pi_{,\n}$. Therefore, we can
redefine the generator $P_\a$,  
\be
P_\a\to\wtilde P_\a\equiv P_\a +E_\a \, ,                    \le{14}
\ee
so that $\tilde P_\a$ has well defined functional derivative. 
The assumed asymptotic behaviour of phase--space
variables ensures {\it finitness} of $E_\a$. 
                 
In a similar way we find $\wtilde P_0\equiv P_0+E_0$, where
\be
E_0\equiv\oint ds_\g(-2aJh_a{^\a}h_b{^\g}A^{ab}{_\a} )\, .    \le{15}
\ee
The surface term $E_0$ is finite under the adopted asymptotic
conditions, and represents the value of the energy of the system.

The spatial rotation generator reads 
$\wtilde M_{\a\b}=M_{\a\b}+E_{\a\b}$, where 
\be
E_{\a\b}\equiv\oint ds_\g \bigl[ -\pi_{\a\b}{^\g}
 +x_{[\a}\bigl( \pi_{ij}{^\g}A^{ij}{_{\b ]}}\bigr) \bigr]\, .\le{16}
\ee
A detailed analysis shows that the adopted asymptotic conditions do not
guarantee the finitness of $E_{\a\b}$, as the
integrand contains $\cO_1$ terms. These troublesome terms are seen to
vanish if we impose the {\it asymptotic\/} gauge condition
$a_{[ij]}=\cO_2$ on the gauge potentials $a^k{_\m}=b^k{_\m}-\d^k_\m$,
and  certain {\it parity conditions\/}. These conditions are invariant
under the global Poincar\'e transformations, and they restrict the
remaining gauge symmetry.  After that $E_{\a\b}$ is
seen to be finite and, consequently, $\wtilde M_{\a\b}$ is well
defined.  

By varying the expression for the boost generator one fins
\be
E_{0\b}\equiv\oint ds_\g \bigl[ -\pi_{0\b}{^\g}
  +x_0 \bigl(\pi_{ij}{^\a}A^{ij}{_{[\b}}\d_{\a ]}{^\g}\bigr) 
  -x_\b \bigl( 2aJh_a{^\a}h_b{^\g}A^{ab}{_\a}\bigr) \bigr]\, . \le{17}
\ee
Additional gauge and parity conditions guarantee the finitness of
$E_{0\b}$. 

All these results are reffered to the EC theory.
Analogous considerations in the general $R+T^2+R^2$ theory show that
the boost generator cannot be redefined by adding a surface term.
Therefore, it is not a well defined generator under the adopted boundary
conditions.
\subsub{Conservation laws.} The improved asymptotic Poincar\'e generators
satisfy the standard Poincar\'e algebra, up to squares (or higher
powers) of constraints and surface terms. This results proves the
asymptotic Poincar\'e symmetry of the theory. We now wish to see
whether this symmetry implies, as usually, the existence of certain
conserved quantities.           

One can prove that a phase-space functional $G[\vphi,\pi,t]$ is a
generator of global symmetries if and only if  
$$
\{ G, \wtilde H_T\} +{\partial G\over\partial t}= C_{PFC} \, ,\qquad
\{ G, \vphi_s\}\approx 0 \, ,                             
$$
where $\wtilde H_T$ is the improved Hamiltonian, $\vphi_s$ are
all constraints, and, as before, the equality means
an equality up to the zero generators.  
The first equation represents the Hamiltonian form of the
conservation law. Indeed, it implies 
${dG /dt}\equiv \{G,H_T\}+\partial G/\partial t\approx S$, so that $G$
is conserved {\it if the surface term $S$ is absent\/}. 

One finds in this way that the generators $\wtilde P_0$, $\wtilde P_\a$ 
and $\wtilde M_{\a\b}$ are conserved, and that the surface
terms $E_0$, $E_\a$ and $E_{\a\b}$ represent the values of energy,
linear momentum and angular momentum as conserved quantities.
On the other hand, the boost generator is not a conserved quantity.
This result is a consequence of an explicit, linear time dependence of
$\wtilde M_{0\b}$, and the existence of a non--vanishing surface term
in $\wtilde P_\b$. 
\subsub{Comparison with the Lagrangian formalism.} In order to
compare the form of the surface terms with those obtained by the
Lagrangian treatment, one should express all momentum variables 
in terms of fields and their derivatives, with the help of the
constraints and the equations of motion.  
One finds that
{~$i)$} the energy--momentum in EC theory is given by the same
expressions as in GR,  
{~$ii)$} the angular momentum also coincides with the GR expression. 

In the general $R+T^2+R^2$ theory, the result for the
energy--momentum is of the same form, while the angular momentum 
remains the same only when all tordions are massive.
When massless tordions exist, then 
$a)$ the spatial angular momentum $E_{\a\b}$ becomes different from the
GR expression, and $b)$ the boost $E_{0\b}$ is not even defined
in this case.

\section{Concluding remarks} 

1) We constructed the Hamiltonian for the general PGT. 
The Hamiltonian constraints $\cH_\bot,\cH_\a,\cH_{ij}$ are 
found to be first class.

2) The Poisson bracket algebra of constraints is calculated and used
to construct the Poincar\'e gauge generators.

3) In case of the Minkowskian asymptotics, we obtained the
conservation of energy--momentum and angular momentum. Other
interesting asymptotic conditions (e.g. de Sitter spacetime) have
not yet been studied.  

4) Depending on the structure of $\tilde\cL$, one may have extra FC
constraints in the theory. The related extra gauge symmetries 
have been studied only in the linear approximation \cite{8}. 

5) The Hamiltonian approach may be very usefull in clarifying the
dynamical structure of the teleparallelism theory. 


\end{document}